\documentclass[journal]{IEEEtran}
\IEEEoverridecommandlockouts
\usepackage{cite}
\usepackage{amsmath,amssymb,amsfonts}
\usepackage{algorithmic}
\usepackage{graphicx}
\usepackage{textcomp}
\usepackage{xcolor}
\usepackage{multirow}
\usepackage{enumitem}
\usepackage{bm}
\usepackage{comment}

\def\BibTeX{{\rm B\kern-.05em{\sc i\kern-.025em b}\kern-.08em
    T\kern-.1667em\lower.7ex\hbox{E}\kern-.125emX}}
\begin{document}

\title{RFSoC-Based Integrated Navigation and Sensing Using NavIC }

\author{
\IEEEauthorblockN{Riya Sachdeva$^\ast$, Aakanksha Tewari$^\ast$, Sumit~J.~Darak,   Shobha~Sundar~Ram, and Sanat K. Biswas}
\IEEEauthorblockA{Indraprashtha Institute of Information Technology Delhi, New Delhi, India \\
E-mail:\{riya22411, aakankshat, sumit, shobha, sanat\}@iiitd.ac.in}
\thanks{*Riya Sachdeva and Aakanksha Tewari are joint first-authors.}
\thanks{This work is supported by the funding received from Chip to Startup (C2S, project no.: EE-9/2/2021-R\&D-E) project from Ministry of Electronics and Information Technology (MeiTy), Government of India.}
}

\IEEEaftertitletext{\vspace{-1.8\baselineskip}}

\maketitle

\begin{abstract}
Prior art has proposed a secondary application for Global Navigation Satellite System (GNSS) infrastructure for remote sensing of ground-based and maritime targets. Here, a passive radar receiver is deployed to detect uncooperative targets on Earth's surface by capturing ground-reflected satellite signals. This work demonstrates a hardware prototype of an L-band Navigation with Indian Constellation (NavIC) satellite-based remote sensing receiver system mounted on an AMD Zynq radio frequency system-on-chip (RFSoC) platform. Two synchronized receiver channels are introduced for capturing the direct signal (DS) from the satellite and ground-reflected signal (GRS) returns from targets. These signals are processed on the ARM processor and field programmable gate array (FPGA) of the RFSoC to generate delay-Doppler maps of the ground-based targets. The performance is first validated in a loop-back configuration of the RFSoC. Next, the DS and GRS signals are emulated by the output from two ports of the Keysight Arbitrary Waveform Generator (AWG) and interfaced with the RFSoC where the signals are subsequently processed to obtain the delay-Doppler maps. The performance is validated for different signal-to-noise ratios (SNR).
\end{abstract}


\vspace{-0.15cm}
\section{Introduction} 
\vspace{-0.05cm}
Global Navigation Satellite System (GNSS) infrastructure is traditionally deployed for positioning, navigation, and timing (PNT) services. Recent advances in integrated navigation and sensing have enabled secondary remote-sensing applications that exploit opportunistic satellite illumination for detecting and tracking uncooperative ground and maritime targets \cite{10726910,10945753}. This application, termed GNSS-reflectometry, is cost-effective, as it promotes spectrum reuse and leverages existing GNSS infrastructure by adding a low-cost passive radar receiver near the observation area. Conventional GNSS remote-sensing hardware testbeds rely on multiple discrete hardware components for RF downconversion and digitization, followed by offline processing in environments such as MATLAB or Python, resulting in higher latency and power consumption \cite{ansari_GPS_SDR_2024,ma2017maritime,ribo2017software}. To address these limitations, this work presents a hardware prototype for passive GNSS-based remote sensing implemented on an AMD radio frequency system-on-chip (RFSoC), offering a compact, scalable, and flexible single-chip platform \cite{10885553}. The hardware prototype is developed for India’s Regional Navigation Satellite System (RNSS), also known as Navigation with Indian Constellation (NavIC) \cite{mruthyunjaya2017irnss}. The use of NavIC for secondary remote-sensing applications remains largely unexplored and is the focus of this work.

\section{NavIC Remote Sensing System}
\begin{figure}[h]
    \centering
    \includegraphics[scale = 0.55]{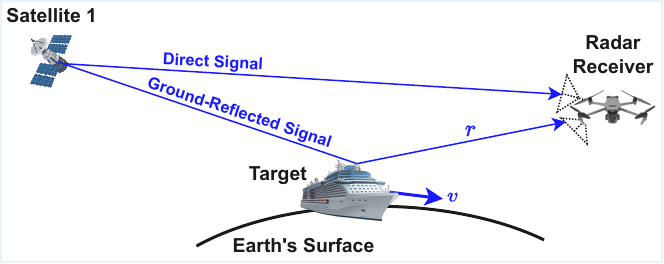}
    \vspace{-0.3cm}
\caption{Bistatic radar system model for NavIC remote sensing}
\label{fig:navic_sysModel}
\end{figure}

We use the NavIC L5 Standard Positioning Service (SPS) centered at $1.176$ GHz. The signal employs a tiered modulation structure with a $1.023$ MHz pseudo-random noise (PRN) sequence as the primary spreading signal to modulate the navigation data. The 1023-bit PRN codes, unique to each satellite and known at the receiver, act as radar waveforms in GNSS-reflectometry due to their orthogonality and excellent autocorrelation properties. The PRN sequence repeats at a 1 ms pulse repetition interval (PRI) with a 100\% duty cycle. The NavIC constellation comprises seven satellites, ensuring illumination of any observation area in India at all times. The remote-sensing system operates as a bistatic radar with the transmitter and receiver at separate locations, as shown in Figure \ref{fig:navic_sysModel}, where the NavIC satellite serves as the transmitter of opportunity. A passive radar receiver mounted on an unmanned aerial vehicle (UAV) is deployed near the observation area to capture the direct signal (DS) from the satellite and the ground-reflected signal (GRS) from the target. The digitized and downconverted received signals on each channel, $y_d$ (DS) and $y_{gr}$ (GRS), are processed separately and modeled as shown, where $k$ is the fast-time index; $k_d$ and $k_{gr}$ are delay indices; $f_d$ and $f_{gr}$ are Doppler shifts; and $A_d$ and $A_{gr}$ denote the amplitudes on the DS and GRS channels respectively. $c_i$ is the PRN-i code corresponding to the $i^{th}$ NavIC satellite with sampling interval $T_s$.
\begin{align}
\tiny
y_{d}[k] =A_{d}c_i[k-k_{d}]e^{j2\pi f_{d}kT_{s}}
\label{eq:los_eq}
\end{align}
\vspace{-0.7cm}
\begin{align}
\tiny
y_{gr}[k] =A_{gr}c_i[k-k_{gr}]e^{j2\pi f_{gr}kT_{s}}
\label{eq:nlos_eq}
\end{align}
\vspace{-0.9cm}
\label{Sec:sys_model}

The digitized raw data on each channel is processed via coarse acquisition (C/A) for satellite identification and delay–Doppler map (DDM) generation. The C/A processing operates on 1 ms of data independently on DS and GRS channels. The DDM involves matched filtering the fast-time samples with the internally generated PRN reference and Doppler processing by multiplying the received samples with candidate Doppler search bins. On the DS channel, the search is performed sequentially for each satellite by generating DDM from cross-correlation with the respective PRN codes. Due to orthogonality, a distinct peak occurs only when the correct PRN codes are cross-correlated. Thus, the transmitting satellite is detected if the DDM peak exceeds a defined threshold. Once identified, DDM generation is carried out for the GRS channel using the selected PRN codes.
\vspace{-0.15cm}
\section{Hardware Setup}
\label{Sec:hw_setup}
We implement the NavIC remote sensing hardware prototype on the AMD Zynq RFSoC $4\times 2$ platform, comprising a quad-core A53 ARM processor (processing system, PS), an Ultrascale FPGA (programmable logic, PL), and integrated on-chip RF data converters (RFDCs). The architecture employs hardware–software co-design, partitioning signal processing tasks between PS and PL. Two synchronized parallel channels are implemented in the PL to capture DS and GRS samples, which are subsequently processed in the PS for DDM generation. Using the PYNQ framework, the DDM and results are displayed in real time. The design is validated over the following two configurations.
\vspace{-0.15cm}
\subsection{RFSoC two-channel loopback NavIC transceiver}
\begin{figure}[h]
    \centering
\includegraphics[scale = 0.55]{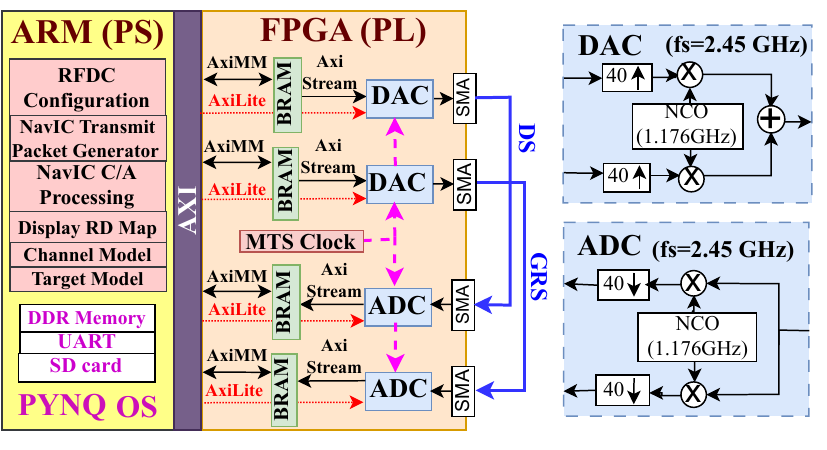}
\vspace{-0.4cm}
\caption{NavIC remote sensing transceiver on RFSoC via two-channel loopback}
\label{fig:hw_arch_loopback}
\end{figure}

The hardware architecture for a two-channel loopback configuration on RFSoC is shown in Figure \ref{fig:hw_arch_loopback}. In PS, the NavIC transmit packet is generated from resampled PRN-2 codes over $1$ ms, upsampled to $61.44$ MHz through a 3-stage soft $\times 2$ interpolation (total $\times 8$). Each interpolation stage has a low-pass FIR filter with 23, 15, and 15 taps, respectively. The channel effects are modeled by adding path loss, time delay corresponding to the propagation path, Doppler shift (from satellite motion), and additive white Gaussian noise (AWGN). Based on the target model, additional delay, Doppler offsets, and amplitude attenuation are incorporated in the GRS. The samples for DS and GRS channels are sent to PL and written to their respective block RAM (BRAM) via the AXI memory-mapped (MM) port. These samples are then sent via 32-bit stream data width at a clock of $61.44$ MHz to the digital-to-analog (DAC) units, configured to accept 2 IQ interleaved samples per clock cycle. All RFDCs operate at $61.44$ MHz. The DAC interpolates the samples by a factor of 40, resulting in a sampling frequency of 2.45 GHz, and upconverts to a center frequency of 1.176 GHz using a numerically controlled oscillator (NCO) via quadrature modulation.

\begin{figure}[h]
\centering
\includegraphics[scale = 0.225]{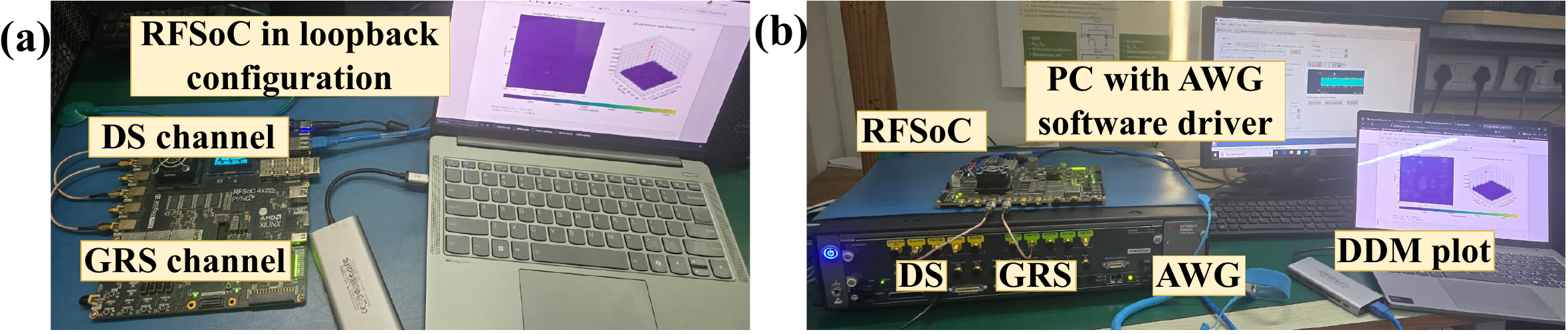}
\vspace{-0.6cm}
\caption{Hardware setup for NavIC remote sensing with (a) Zynq RFSoC in loopback configuration, (b) Keysight AWG generating the NavIC waveform being fed into the RFSoC-based NavIC receiver.}
\label{fig:hw_setup}
\end{figure}

Figure \ref{fig:hw_setup}(a) shows the hardware setup for the two-channel RFSoC loopback. Here, the DS and GRS signals are converted to analog and transmitted to the corresponding ADC via a two-channel wired loopback. At the ADC, the signals are sampled at $2.45$ GHz, downconverted to baseband using quadrature demodulation, and downsampled by a factor of 40 to $61.44$ MHz. The IQ samples are sent from ADC to BRAM via separate stream interfaces, each with a data width of 16 bits at a rate of 1 sample/cycle. They are sent to PS via the AXI-MM interface for 3-stage soft decimation by 2 (total 8) to a sampling rate of 7.68 MHz. This is followed by NavIC C/A processing for DDM generation, involving cross-correlation with resampled PRN codes on each channel. This architecture requires four parallel processing chains on the FPGA, with each RFDC on a separate tile. Multi-tile synchronization (MTS) is enabled between DACs and ADCs using a clock generated from an internal source with reference from the external LMX2594 chip. 

\vspace{-0.3cm}

\vspace{-0.5cm}
\vspace{-0.4cm}
\subsection{AWG-based NavIC simulator with RFSoC-based receiver}

\begin{table}[!t]
\caption{Target estimates with RFSoC loopback and RFSoC+AWG configurations.}
\label{tab:results}
\centering

\fontsize{6}{6}\selectfont                 
\renewcommand{\arraystretch}{1.3}  
\setlength{\tabcolsep}{3pt}        

\begin{tabular}{|c|ccc|ccc|ccc|}
\hline
\multirow{2}{*}{\textbf{\begin{tabular}[c]{@{}c@{}}SNR\\  (dB)\end{tabular}}} &
  \multicolumn{3}{c|}{\textbf{Ground Truth}} &
  \multicolumn{3}{c|}{\textbf{RFSoC Loopback Estimates}} &
  \multicolumn{3}{c|}{\textbf{RFSoC+AWG Estimates}} \\ \cline{2-10} 
 &
  \multicolumn{1}{c|}{\textbf{\begin{tabular}[c]{@{}c@{}}Range \\ Offset \\ (km)\end{tabular}}} &
  \multicolumn{1}{c|}{\textbf{\begin{tabular}[c]{@{}c@{}}Doppler \\ DS \\ (Hz)\end{tabular}}} &
  \textbf{\begin{tabular}[c]{@{}c@{}}Doppler\\ GRS \\ (Hz)\end{tabular}} &
  \multicolumn{1}{c|}{\textbf{\begin{tabular}[c]{@{}c@{}}Range \\ Offset \\ (km)\end{tabular}}} &
  \multicolumn{1}{c|}{\textbf{\begin{tabular}[c]{@{}c@{}}Doppler \\ DS \\ (Hz)\end{tabular}}} &
  \textbf{\begin{tabular}[c]{@{}c@{}}Doppler \\ GRS \\ (Hz)\end{tabular}} &
  \multicolumn{1}{c|}{\textbf{\begin{tabular}[c]{@{}c@{}}Range \\ Offset \\ (km)\end{tabular}}} &
  \multicolumn{1}{c|}{\textbf{\begin{tabular}[c]{@{}c@{}}Doppler \\ DS \\ (Hz)\end{tabular}}} &
  \textbf{\begin{tabular}[c]{@{}c@{}}Doppler \\ GRS \\ (Hz)\end{tabular}} \\ \hline
-5 &
  \multicolumn{1}{c|}{4} &
  \multicolumn{1}{c|}{1500} &
  500 &
  \multicolumn{1}{c|}{3.984} &
  \multicolumn{1}{c|}{1500} &
  500 &
  \multicolumn{1}{c|}{4.023} &
  \multicolumn{1}{c|}{1500} &
  500 \\ \hline
-10 &
  \multicolumn{1}{c|}{6} &
  \multicolumn{1}{c|}{1000} &
  -500 &
  \multicolumn{1}{c|}{6.015} &
  \multicolumn{1}{c|}{1000} &
  -500 &
  \multicolumn{1}{c|}{6.015} &
  \multicolumn{1}{c|}{1000} &
  -500 \\ \hline
-12 &
  \multicolumn{1}{c|}{8} &
  \multicolumn{1}{c|}{500} &
  -500 &
  \multicolumn{1}{c|}{8.007} &
  \multicolumn{1}{c|}{500} &
-500 &
  \multicolumn{1}{c|}{8.007} &
  \multicolumn{1}{c|}{500} &
  -500 \\ \hline
\end{tabular}

\vspace{-0.5cm}

\end{table}

\begin{figure}[h]
    \centering
\includegraphics[scale = 0.5]{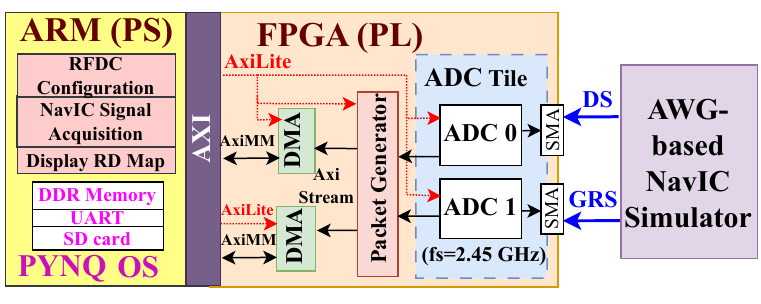}
\vspace{-0.4cm}
\caption{NavIC remote sensing two-channel receiver on Zynq RFSoC interfaced with AWG-based \vspace{-0.5cm}}
\label{fig:hw_arch_awg}
\end{figure}

Here, we replace the RFSoC-based NavIC transmitter with a Keysight M8910A arbitrary waveform generator (AWG) to emulate DS and GRS NavIC signals at RF. Figure \ref{fig:hw_setup}(b) shows the hardware setup of this configuration. The samples corresponding to the NavIC signal arriving at the receiver (upconverted to $1.176$ GHz) with channel effects and target returns are loaded into the AWG with channel coupling enabled from the driver software. The AWG uses 12-bit ADCs clocked at $2.45$ GHz to output the DS and GRS signals on two synchronized channels. The RF signal is fed into the RFSoC, where receiver C/A processing takes place after digitization by ADCs. The RFSoC hardware architecture for the NavIC receiver is shown in Figure \ref{fig:hw_arch_awg}, illustrating capture of DS and GRS samples via two ADCs. Since the RFSoC $4\times2$ supports two ADC units on the same tile, MTS is not required. The ADCs are configured with identical settings as discussed in Section \ref{Sec:hw_setup}. The samples from the ADC are streamed to a packet generator IP core, which groups $1$ ms of samples and sends the DS and GRS packets to PS over two DMA cores for NavIC C/A processing.
\vspace{-0.3cm}
\vspace{-0.1cm}
\section{Results and Discussions}
\begin{figure}[h]
    \centering    \includegraphics[scale = 0.15]{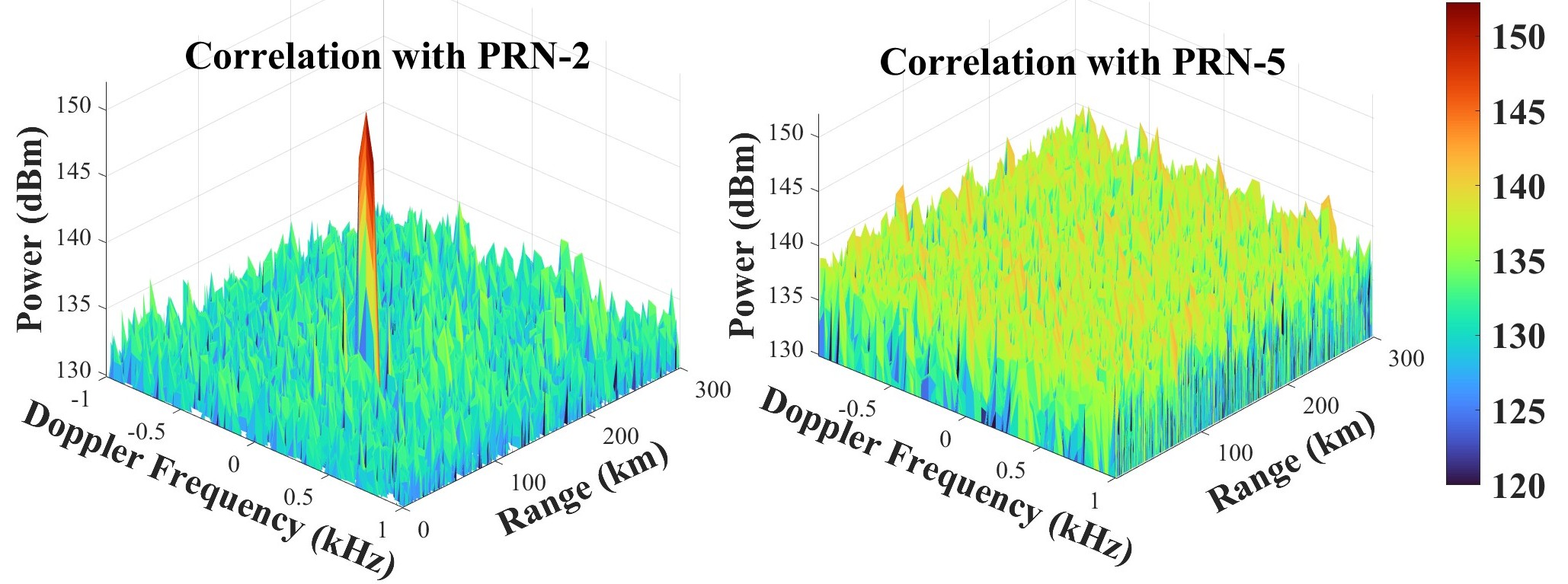}
\caption{ Satellite detection DDMs for NavIC signal coming from PRN-2.}
\label{fig:sat_detection}
\end{figure}

\begin{figure}[h]
    \centering
    \includegraphics[scale = 0.25]{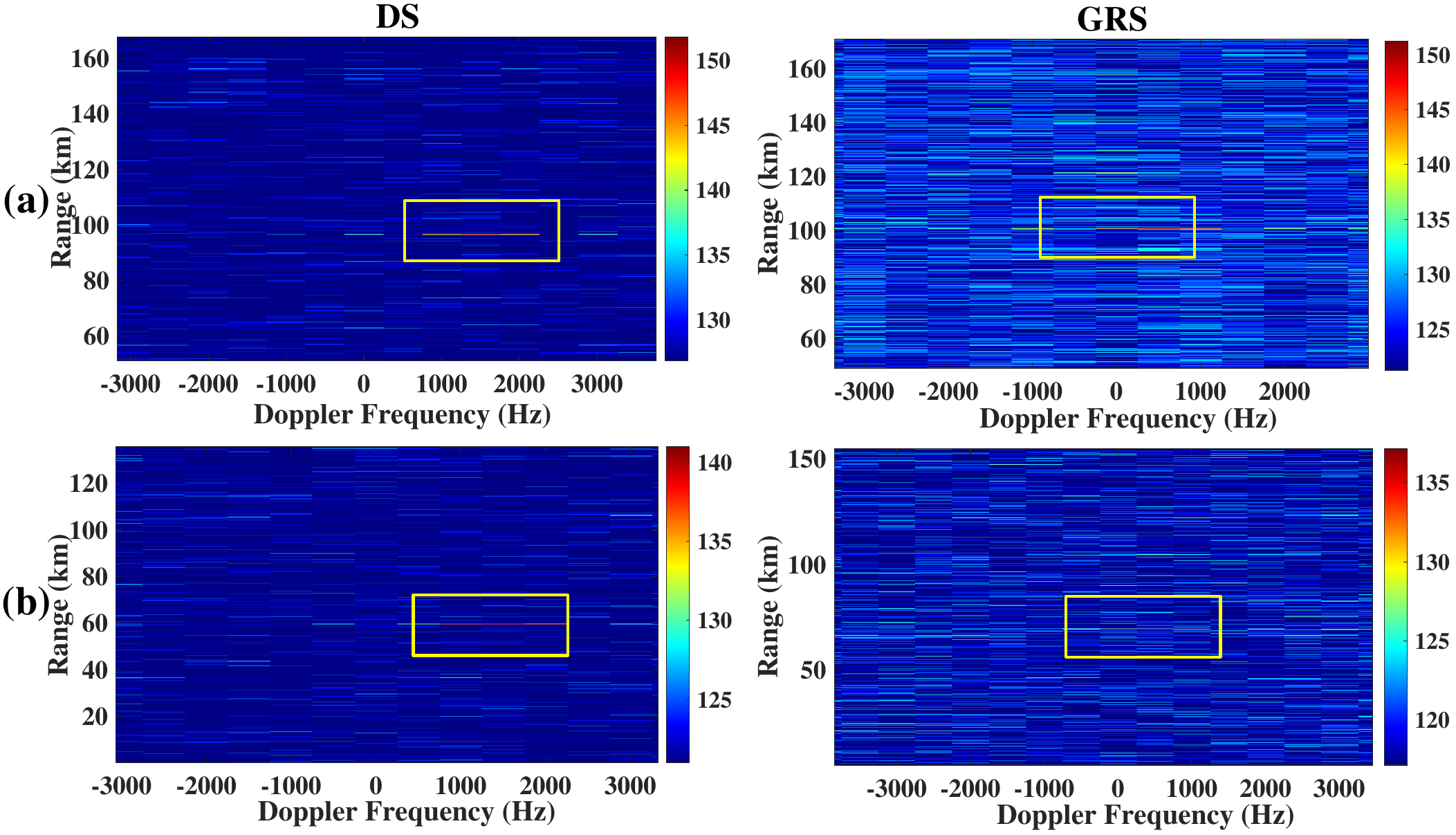}
\caption{DDM with (a) RFSoC loopback and (b) AWG+RFSoC setup for received SNR of -10 dB \vspace{-0.8cm}}
\label{fig:ddm}
\end{figure}

We characterize the NavIC remote sensing performance for the two hardware configurations discussed in Section \ref{Sec:hw_setup} in terms of target range and Doppler detection metrics obtained from DDM. We consider the NavIC satellite with PRN-2 illuminating the observation area and a point target located within a $10$ km radius from the stationary radar receiver. The NavIC L-band signal corresponds to a bandwidth of $2$ MHz, offering a range resolution of $293$ m. We oversample the signal at the receiver to $7.68$ MHz to obtain a finer precision of $40$ m. The C/A is performed across 41 Doppler bins spanning -10 to 10 kHz at $500$ Hz intervals. The signal-to-noise ratio (SNR) is varied from -5 to -12 dB by artificially changing the noise floor by adding AWGN noise to the transmitted signal. Figure \ref{fig:sat_detection} shows the DDMs for satellite detection on the DS channel at an SNR of -10 dB from the RFSoC loopback configuration. It indicates that cross-correlation with the PRN-2 reference sequence yields a distinct peak with a gain of up to 20 dB compared to other satellites (PRN-5), thus indicating the presence of satellite 2 in the signal. Figure \ref{fig:ddm} shows the DS and GRS DDM from RFSoC loopback and AWG+RFSoC configurations, both successfully detecting targets spaced $4$ km apart with Doppler returns of $500$ Hz under an SNR of -5 dB. Here, the bistatic range of the target is calculated from the delay offset $k_{gr}-k_d$, where $k_d$ and $k_{gr}$ are obtained from the peaks of the DS and GRS DDMs, respectively. The bistatic Doppler, $f_{gr}$, is obtained from the GRS channel. While both channels can perform successful detection up to an SNR of -12 dB, the post-processing gain on the GRS channel is lower than that of DS due to the weaker GRS signal after target reflection. Table \ref{tab:results} presents the target detections from the two hardware configurations for various transmitter and target parameters and different SNR levels, benchmarked against the ground truth values. {At -5 dB SNR, the average RMSE in bistatic range and Doppler is 0.14 km and 250 Hz, respectively, both within the corresponding resolution limits. The estimation precision can be further improved by increasing the FFT size for frequency-domain range processing and using finer Doppler search-bin spacing.} 

{The work in \cite{ansari_GPS_SDR_2024} presents a Nuand BladeRF-based GNSS receiver, while \cite{ribo2017software} discusses a software interferometric receiver; both support only offline processing and do not enable real-time analysis. In contrast, the proposed RFSoC-based solution achieves an execution time of $274$ ms for the complete NavIC receiver chain at $3.75$ W power consumption. The C/A processing is performed in the PS, and offloading it to the PL provides an additional $3\times$ acceleration.  }
\label{Sec:results}
\vspace{-0.5cm}

\section{Conclusion and Future Works}
We implemented an RFSoC-based hardware prototype for NavIC remote sensing with two synchronized receiver channels for processing DS and GRS signals. DDM generation was validated in RFSoC loopback and AWG+RFSoC configurations. The current hardware setup does not incorporate link-budget analysis for the highly attenuated satellite signals at the radar receiver. Future work will extend the prototype to capture real NavIC signals using an analog front end (AFE).
\label{Sec:conclusion}

\bibliographystyle{IEEEtran}
\bibliography{References}
\end{document}